\documentclass[11pt,a4paper]{article}
 \usepackage{jheppub}
 \usepackage{axodraw}
 \newcommand{\jouref}[4]{{#1 }{\bf #2} (#3) #4}
 
 \newcommand{\hepph}[1]{{hep-ph/#1}}

 \title{Standard Model gauge couplings from gauge--dilatation symmetry 
breaking}

 \author{Kosuke Odagiri}
 \affiliation{Electronics and Photonics Research Institute,
 National Institute of Advanced Industrial Science and Technology,
 Tsukuba Central 2,
 1--1--1 Umezono, Tsukuba, Ibaraki 305--8568, Japan}

 \abstract{
  We argue that there is a spontaneously broken rotational symmetry 
between space-time coordinates and gauge theoretical phases.
  The dilatonic mode acts as the massive Higgs boson, whose vacuum 
expectation value determines the gauge couplings.

  This mechanism requires that the quadratic divergences, or tadpoles of 
the three gauge-theory couplings, unify at a certain scale.
  We verify this statement, and find that this occurs at 
$\Lambda_\mathrm{u}\approx4\times10^7$~GeV.

  The tadpole cancellation condition, together with the dilaton 
self-energy, fixes the value of the unified tadpole coefficient to be 
$\left[4\ln(\Lambda_\mathrm{cut}/\Lambda_\mathrm{u})\right]^{-1}$.
  The observed values of the coupling constants at $\Lambda_\mathrm{u}$ 
then implies $\Lambda_\mathrm{cut}\approx4\times10^{18}$~GeV, which is 
close to the value of the reduced Planck mass 
$\widetilde{M}_\mathrm{Pl}=M_\mathrm{Pl}/\sqrt{8\pi}=2.4\times10^{18}$~GeV.
  In other words, by assuming a cutoff at $M_\mathrm{Pl}$ or 
$\widetilde{M}_\mathrm{Pl}$, we are able to obtain predictions for the 
gauge couplings which agree with the true values to within a few 
percent.

  It turns out that this symmetry breaking can only take place if mass 
is generated with the aid of some other means such as electroweak 
symmetry breaking.
  Assuming dynamical symmetry breaking originating at 
$\widetilde{M}_\mathrm{Pl}$, we obtain $M_\chi\approx10^9$~GeV, which is 
not unreasonable but somewhat higher than $\Lambda_\mathrm{u}$.

  The cancellation of an anomaly in the dilaton self-energy requires 
that the number of fermionic generations equals three.
 }

\begin{document}

 \maketitle

 \section{Introduction}

  One of the fundamental questions of particle physics is that of what 
determines the gauge couplings, or indeed of what gauge bosons are.

  In this paper, we address these questions by extending the Standard 
Model to a theory in which the gauge bosons arise as effective 
Nambu--Goldstone modes (denoted Goldstone modes hereafter) of a 
spontaneously broken symmetry.
  Provided that the symmetry is broken dynamically 
\cite{nambu,bardeen,topcondensation}, we can calculate the parameters of 
the theory by requiring self-consistency.

  What can be this broken symmetry?

  Goldstone modes in general have the form of the rotation operators of 
the corresponding broken symmetry, so we ask which rotation is generated 
by the gauge-field operator $A^i_\mu$, and we come to the conclusion 
that the symmetry is between gauge-theoretical phases and space--time 
coordinates.
  This is not unlike Kaluza--Klein theory \cite{kaluza-klein}.
  However, the structure of space-time itself is not affected in our 
analysis.
  Gravity is small and negligible here, and is important only for 
setting the cutoff scale.

  Whenever a symmetry is broken spontaneously, there arises a Goldstone 
mode for each broken symmetry and a Higgs mode for a preserved symmetry.
  The preserved symmetry in this case is the approximate scale 
invariance, or dilatation.

  Note that the dilaton in the context of our study is not so much a 
pseudo-Goldstone mode for broken scaling symmetry \cite{coleman} as a 
massive Higgs mode corresponding, somewhat paradoxically, to the 
preserved scaling symmetry.

  The gauge coupling constants are proportional to the inverse of the 
vacuum expectation value $v_\chi$ of this dilaton.

  In this paper, we calculate $v_\chi$ and the dilaton mass $M_\chi$
within the context of dynamical symmetry breaking.
  These are calculable because $v_\chi^2$ is proportional to the dilaton 
self-energy whereas $M_\chi$ is fixed by the cancellation condition of 
the gauge-boson tadpoles.
  The ratio of the two quantities is more easily calculable than the two 
quantities separately. This ratio gives us the gauge coupling constants 
at the $M_\chi$ scale.

  In order that this mechanism works, it is necessary that the three 
tadpoles unify at a scale $\Lambda_\mathrm{u}\approx M_\chi$.
  This is verified phenomenologically. The unification scale turns out 
to be $\Lambda_\mathrm{u}\approx4\times10^7$~GeV, including 
next-to-leading-order running effects.

  Our procedure utilizes an analogous framework that was developed in 
refs.~\cite{gribovlongshort,gribovquarkconfinement} within the context 
of chiral symmetry breaking and 
refs.~\cite{gribovewsb,odagirimagnetism,dasodagiri} for the case of the 
Higgs mechanism.

  Let us denote the gauge theoretical tadpole coefficients as $c_i$ 
($i=1,2,3$), where the tadpole is given as a function of the cutoff 
scale $\Lambda$ by $\alpha_ic_i\Lambda^2$.
  Our prediction for the gauge couplings is then
 \begin{equation}
  (c_i\alpha_i)^{-1}=4\log(\Lambda_\mathrm{cut}/\Lambda_\mathrm{u}).
 \end{equation}
  A natural guess for $\Lambda_\mathrm{cut}$ would be 
$M_\mathrm{Pl}=\sqrt{\hbar c/G_\mathrm{N}}$ or 
$\widetilde{M}_\mathrm{Pl}=\sqrt{\hbar c/8\pi G_\mathrm{N}}$.
  Adopting $\Lambda_\mathrm{cut}=\widetilde{M}_\mathrm{Pl}$ with one 
order-of-magnitude error estimation on each side, we then obtain
 \begin{equation}
  (c_i\alpha_i)^{-1}=99\pm9.
 \end{equation}
  The phenomenological value turns out to be
 \begin{equation}
  (c_i\alpha_i)^{-1}=101.9,
 \end{equation}
  in good agreement with the prediction.

  The disparity between $\Lambda_\mathrm{cut}$ and $\Lambda_\mathrm{u}$ 
requires explanation.

  In order that the symmetry-broken vacuum is stable, it is necessary 
that the tadpole of the order parameter, i.e., the dilaton, vanishes.
  It turns out that if we consider the gauge--dilatation symmetry 
breaking alone, we can never satisfy this condition, because there is no 
term that cancels the dilaton self-coupling term.
  Some other mass-generation mechanism is necessary.
  The obvious choice would be electroweak symmetry breaking (EWSB).

  If the masses of all Standard Model particles are due to the Higgs 
mechanism, only the Higgs-boson loop contributes to the dilaton tadpole.
  The tadpole cancellation condition is of the form 
$\Lambda_\mathrm{cut}M_\mathrm{Higgs}\sim M_\chi^2$, and therefore the 
disparity is explained.
  However, a more detailed calculation based on the assumption of 
dynamical EWSB at $\Lambda_\mathrm{cut}=\widetilde{M}_\mathrm{Pl}$ 
yields $M_\chi=10^9$~GeV, which is one order-of-magnitude larger than 
the phenomenological value of $\Lambda_\mathrm{u}$.
  Further investigation into this point will be necessary.

  This paper is organized as follows.
  We write down the effective Lagrangian in 
section~\ref{sec_lagrangian}.
  We first discuss the gauge-theory tadpoles phenomenologically in 
section~\ref{sec_gauge_tad}.
  The parameters of the Lagrangian are worked out in the context of 
dynamical symmetry breaking in section~\ref{sec_derivation}.
  The paper is concluded with a summary and brief discussions on future 
extensions.

 \section{The Lagrangian}
 \label{sec_lagrangian}

  As an introductory remark, when a symmetry is broken spontaneously in 
general, a Higgs mode and Goldstone modes arise, and they have the 
following properties:
 \begin{enumerate}
  \item The Goldstone modes are massless (gapless), and their coupling 
has the form of the broken symmetry operation.
  This form of coupling is necessary if the Goldstone boson corresponds 
to the broken part of the of the symmetry current.
  \item The number of Goldstone modes equals the number of broken 
symmetries.
  \item The Goldstone and Higgs fields themselves are connected by the 
same rotation symmetry.
  \item The Higgs mode is massive (finite energy gap), and their 
coupling has the form of the preserved symmetry operation. This form of 
coupling is required by the symmetry between Goldstone and Higgs fields.
  \item The form factor for Goldstone modes, i.e., the inverse of their 
coupling strengths, is proportional to the vacuum expectation value 
(VEV) of the Higgs field.
 \end{enumerate}

  Our proposal is that the gauge bosons are the Goldstone modes of a 
certain symmetry.
  If so, the theory which satisfies the above properties can be written 
down almost uniquely (unique up to the kinetic terms and the potential).
  To start off with, from the condition that the vector field $A_\mu^i$ 
behaves like the rotation operation, we conclude that the symmetry is 
between gauge-theoretical phases and space-time coordinates.

  The four-vector $dx^\mu$ is thus generalized to
 \begin{equation}
  (dx^\mu,\ r_0d\phi^i).
  \label{eqn_extended_dx}
 \end{equation}
  Here, $\phi^i$ are the gauge-theoretical phases. $i$ is the index for 
the generators, e.g.\ 1 to 8 for QCD.
  The generalization to multiple gauge groups is trivial.
  $r_0$ is some number with the dimension of length.
  This is proportional to the coupling constants $g$.

\begin{figure}[ht]
  \centerline{
  \begin{picture}(200,100)(0,0)
  \put(30,30){\includegraphics[width=4.5cm]{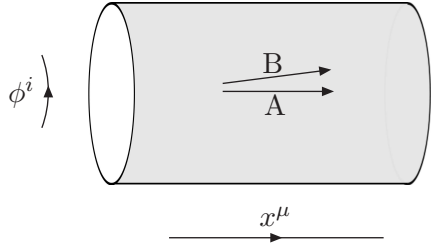}}
  \LongArrow(80,65)(120,65)
  \LongArrow(80,68)(119,73)
  \Text(100,60)[c]{A}
  \Text(99,76)[c]{B}
  \ArrowLine(60,10)(140,10)
  \Text(100,18)[c]{$x^\mu$}
  \ArrowArc(-25,65)(40,-20,20)
  \Text(5,65)[c]{$\phi^i$}
  \end{picture}
  }
  \caption{\label{fig_cylinder}
  An illustration based on Kaluza--Klein-like picture.
  Gauge phase $\phi^i$ and space-time coordinates $x^\mu$ mix.
  A rotation in $(x,\phi)$ space maps, for example, vector $A$ to vector 
$B$.}
 \end{figure}

  In Kaluza--Klein theories \cite{kaluza-klein}, $r_0$ corresponds to 
the radius of the compactified dimension.
  In our picture, $r_0$ is a parameter with no geometrical significance 
and is fixed dynamically as a function of the energy scale.
  Even so, the picture of compactified dimensions, depicted in 
figure~\ref{fig_cylinder}, is a useful aid.
  In terms of figure~\ref{fig_cylinder}, whereas Kaluza--Klein theory 
deals with oscillations of the cylinder itself, we are dealing with the 
symmetry operations on the `mesh' that may be imagined to be drawn on 
the cylinder.

  In Kaluza--Klein theory, one starts from a $4+\mathrm{n}$ dimensional 
theory and then apply compactification to obtain the effective theory in 
$4$ dimensions.
  Our approach differs in that we do not start from a $4+\mathrm{n}$ 
dimensional theory.
  The theory is defined only in $4$ dimensions, and the parameters of 
the theory are worked out from the condition of self-consistency as 
applied to the $4$ dimensional theory, which arises because of the 
spontaneous symmetry breaking.

  $x^\mu$ and $\phi^i$ are quite distinct entities even though the 
rotation between them is a symmetry operation.
  $\phi^i$ correspond to extra-dimensional rotation angles rather than 
independent dimensions.
  As an illustration, take the example of the n-sphere as compactified 
dimensions.
  The number of coordinates is equal to n, whereas the number of 
rotations among these coordinates is equal to 
$\frac12\mathrm{n}(\mathrm{n}+1)$, i.e., equal to the order of the 
rotational group SO($\mathrm{n}+1$).


  The proper time interval for eqn.~(\ref{eqn_extended_dx}) is given by
 \begin{equation}
  d\tau^2=dx^2-r_0^2(d\phi^i)^2.
 \end{equation}
  The space-time derivative is replaced in the UV by
 \begin{equation}
  \frac{\partial}{\partial x^\mu}\to
  \left(\frac{\partial}{\partial x^\mu},\
  \frac1{r_0}\frac{\partial}{\partial\phi^i}\right).
 \end{equation}

  The rotation between gauge rotation and space-time translation is 
given by the following transformation, which preserves $d\tau$:
 \begin{equation}
  \left(
  \begin{array}{c} dx^\mu\\ r_0d\phi^i \end{array}
  \right)\longrightarrow
  (1-r_0^2\hat{A}^2)^{-\frac12}
  \left(
  \begin{array}{c}
  dx^\mu+r_0^2\hat{A}^\mu_id\phi^i\\
  r_0d\phi^i+r_0\hat{A}^i\cdot dx
  \end{array}
  \right).
  \label{eqn_gauge_spacetime_rotation}
 \end{equation}
  This is just rotation in 
$1_\mathrm{time}+(3+n)_\mathrm{space}$-dimensional space (see 
figure~\ref{fig_cylinder}) where $n$ is the number of generators.
  The spontaneous breaking of this symmetry implies that the vacuum 
chooses an arbitrary $1+3$-dimensional time-space direction out of the 
$1+(3+n)$ dimensions.
  The space-time metric is given by $\eta^{\mu\nu}$.

  The preserved symmetry operation is dilatation in the $1+3$ 
space-time.
  Including dilatation, the above rotation operation can be cast into 
the following form:
 \begin{equation}
  \left(
  \begin{array}{c} dx^\mu\\ r_0d\phi^i \end{array}
  \right)\longrightarrow
  \left(
  \begin{array}{cc} (v_\chi+\chi){\eta^\mu}_\nu & A_j^\mu\\
  A^i_\nu & {\Phi^i}_j\end{array}
  \right)
  \left(
  \begin{array}{c} dx^\nu\\ r_0d\phi^j \end{array}
  \right).
  \label{eqn_extended_vierbein}
 \end{equation}
  This time, $d\tau$ is not necessarily preserved.
  We then identify $\chi$ with the dilaton ($v_\chi$ is the vacuum 
expectation value) and $A_\mu^i$ with the gauge fields.
  $\Phi$ do not become physical modes in our situation because they are 
not involved in symmetry breaking.

  We should clarify in which sense the $A_\mu^i$ field behaves as a 
Goldstone mode.
  Goldstone bosons are divergences of a Noether current when the 
symmetry is spontaneously broken.
  The Noether current is given, as usual, by the derivative of the 
Lagrangian with respect to the derivative of the symmetry 
transformation.
  That is,
 \begin{equation}
  J^{\mu\nu}_i=\frac{\partial{\mathcal{L}}}{
  \partial\left(\partial a^i_\nu/\partial x^\mu\right)}.
  \label{eqn_unmodified_current}
 \end{equation}
  $a^i_\nu$ stands for the rotation between $dx^\mu$ and $d\phi^i$, or 
between $(v_\chi+\chi){\eta^\mu}_\nu$ and $A_\nu^i$.
  Note that the two Lorentzian indices $\mu$ and $\nu$ are inequivalent.

  We must write the (4-dimensional) Lagrangian in a way that respects 
the symmetry between the dilaton and the gauge fields.
  This may be achieved by replacing space-time derivatives with
 \begin{equation}
  D_\mu=(v_\chi+\chi){\eta_\mu}^\nu\frac{\partial}{\partial x_\nu}
  -A_\mu^i\frac{\partial}{r_0\partial\phi^i}.
  \label{eqn_dmu}
 \end{equation}
  The negative sign arises because $\phi^i$ are space-like.

  The $\phi^i$ derivative operator is such that
 \begin{eqnarray}
  \frac{\partial}{\partial\phi^i}\psi&=&-i\tau_i\psi,\\
  \frac{\partial}{\partial\phi^i}A^j &=&-f_{ijk}A^k,\\
  \frac{\partial}{\partial\phi^i}\chi&=&0.
  \label{eqn_phi_derivative}
 \end{eqnarray}
  $\tau$ and $f_{ijk}$ are (for QCD) the colour matrices.
  Note that the definition of $D_\mu$ is consistent with the usual 
gauge-theoretical covariant derivative.

  This formulation is not without problems.
  When we try to define $F_{\mu\nu}$ as the commutator of covariant 
derivatives, we find that that there is a non-factorizable contribution:
 \begin{equation}
  (v_\chi+\chi)\frac{\partial\chi}{\partial x_\mu}
  \frac{\partial\psi}{\partial x_\nu}-
  (v_\chi+\chi)\frac{\partial\chi}{\partial x_\nu}
  \frac{\partial\psi}{\partial x_\mu}.
 \end{equation}
  This will vanish in the limit of soft $\chi$, in which case we may 
write
 \begin{equation}
  F_{\mu\nu}=(v_\chi+\chi)\left(
  \frac{\partial A_\nu^i}{\partial x^\nu}-
  \frac{\partial A_\mu^i}{\partial x^\nu}\right)-
  \frac1{r_0}{f^i}_{jk}A_\mu^jA_\nu^k.
 \end{equation}
  In principle, there will be problems with gauge invariance when $\chi$ 
is not soft.

  Up to the normalization factors, the symmetry-conserved part of the 
Lagrangian is then written down trivially as
 \begin{equation}
  \overline{\psi}i\gamma^\mu D_\mu\psi-
  \frac14F_{\mu\nu}F^{\mu\nu}
  +\frac32(v_\chi+\chi)^2
  \left(\frac{\partial\chi}{\partial x_\mu}\right)^2.
  \label{eqn_lagrangian_initial}
 \end{equation}
  We used the third equation of eqns.~(\ref{eqn_phi_derivative}) in the 
last term.
  The action $S$ is $i$ times the four-dimensional integral of 
$\mathcal{L}$.
  This Lagrangian is quite different from interactions that involve 
the gravitational dilaton (for standard review papers, see 
refs.~\cite{grav_grw,grav_hlz}).

  When $v_\chi=0$, no equation of motion arises for any field, unlike 
any other instances of spontaneous symmetry breaking that we know of.

  As we are only interested in the case $v_\chi\ne0$, let us redefine 
$\psi$ and $A$ so as to absorb $v_\chi$.
  We also normalize $v_\chi$ and $\chi$ such that they have mass 
dimension 1.
  The $\phi^i$ derivative may be replaced by 
eqns.~(\ref{eqn_phi_derivative}).
  The dimensionless quantity $(r_0v_\chi)^{-1}$ can be replaced with the 
gauge coupling strength $g$.

  For the fermionic part, the Lagrangian with the appropriate 
normalization factors is then given by
 \begin{equation}
  \mathcal{L}_\mathrm{f}=
  \overline{\psi}i\gamma^\mu\left[
  \left(1+\frac{\chi}{v_\chi}\right)\frac{\partial}{\partial x^\mu}+
  igA_\mu^i\tau_i
  \right]\psi.
  \label{eqn_fermionic_lagrangian}
 \end{equation}

  The bosonic part is written as
 \begin{equation}
  \mathcal{L}_\mathrm{b}=T_A+T_\chi-V(A,\chi),
 \end{equation}
  where
 \begin{equation}
  T_A=-\frac14\left[
  \left(1+\frac{\chi}{v_\chi}\right)
  \left(\frac{\partial A^i_\nu}{\partial x^\mu}-
  \frac{\partial A^i_\mu}{\partial x^\nu}\right)-
  g{f^i}_{jk}A_\mu^jA_\nu^k\right]^2,
 \end{equation}
 \begin{equation}
  T_\chi=\frac32(1+\chi/v_\chi)^2
  \left(\frac{\partial\chi}{\partial x^\mu}\right)^2,
 \end{equation}
  and the symmetry breaking potential is given by
 \begin{equation}
  V(A,\chi)=\frac{\mu_\chi^2}{8v_\chi^2}
  \left[4\left(v_\chi+\chi\right)^2-\left(A_\mu^i\right)^2
  -4v_\chi^2\right]^2.
 \end{equation}
  Factors of $4$ inside square brackets are due to the trace of 
$\eta_{\mu\nu}$.
  Obviously $16\mu_\chi^2=3M_\chi^2$.
  It is worth pointing out here that effective $A^4$ contact terms, 
which are proportional to $\mu_\chi^2/v_\chi^2$, cancel in the 
low-energy limit $Q^2\ll \mu_\chi^2$.

  Gauge fixing is problematic.
  Ordinarily, we enforce transversality by bringing in covariant gauge 
fixing terms of the form
 \begin{equation}
  -\frac1{2\lambda}\left(
  \frac{\partial A_\mu^i}{\partial x_\mu}\right)^2+
  (\mbox{ghost term}).
 \end{equation}
  However, this does not quite work because the potential $V(A,\chi)$ 
violates gauge symmetry at scales that are comparable with $M_\chi$.
  The resolution of this problem requires a study into the structure of 
$M_\chi$. 
  Our strategy is to relate $M_\chi$ to the gauge theoretical tadpoles.
  If calculations are done self-consistently, this implies that 
longitudinal components do cancel.
  That is, in practical terms, we may adopt the Feynman gauge 
$\lambda=1$.
  This point will be discussed in more detail in 
section~\ref{sec_dilatonic_mass_behaviour}.

\begin{figure}[ht]

  \begin{center}

  \begin{picture}(100,100)(0,0)
   \DashLine(20,60)(80,60){3}
   \LongArrow(40,80)(60,80)
   \Text(50,90)[c]{$k_\mu$}
   \Text(50,20)[c]{$i/(3k^2-16\mu_\chi^2+i0)$}
  \end{picture}
  \begin{picture}(100,100)(0,0)
   \DashLine(20,70)(55,70){3}
   \ArrowLine(80,45)(55,70)
   \ArrowLine(55,70)(80,95)
   \LongArrow(55,80)(70,95)
   \LongArrow(70,45)(55,60)
   \Text(58,45)[c]{$k_1$}
   \Text(58,95)[c]{$k_2$}
   \Text(50,20)[c]{$\frac12iv_\chi^{-1}(\not\! k_1+\not\! k_2)$}
  \end{picture}
  \begin{picture}(100,100)(0,0)
   \DashLine(20,90)(50,75){3}
   \DashLine(50,45)(50,75){3}
   \DashLine(80,90)(50,75){3}
   \Text(50,28)[c]{$3iv_\chi^{-1}\bigl(-16\mu_\chi^2$}
   \Text(50,10)[c]{$+k_1^2+k_2^2+k_3^2\bigr)$}
  \end{picture}
  \begin{picture}(100,100)(0,0)
   \DashLine(20,90)(50,70){3}
   \DashLine(80,90)(50,70){3}
   \DashLine(20,50)(50,70){3}
   \DashLine(80,50)(50,70){3}
   \Text(50,28)[c]{$6iv_\chi^{-2}\bigl(-8\mu_\chi^2+$}
   \Text(50,10)[c]{$k_1^2+k_2^2+k_3^2+k_4^2\bigr)$}
  \end{picture}

  \vspace{7mm}

  \begin{picture}(100,100)(0,0)
   \Photon(20,90)(50,75){-3}{3.5}
   \Text(10,90)[c]{$\mu$}
   \LongArrow(40,70)(20,80)
   \Text(25,70)[c]{$k_1$}
   \Photon(80,90)(50,75){3}{3.5}
   \Text(90,90)[c]{$\nu$}
   \LongArrow(60,70)(80,80)
   \Text(75,70)[c]{$k_2$}
   \DashLine(50,45)(50,75){3}
   \Text(50,28)[c]{$2iv_\chi^{-1}\bigl(2\mu_\chi^2\eta^{\mu\nu}+$}
   \Text(50,10)[c]{$k_1\cdot k_2\eta^{\mu\nu}
   -k_1^\nu k_2^\mu\bigr)$}
  \end{picture}
  \begin{picture}(100,100)(0,0)
   \Photon(20,90)(50,70){-3}{3.5}
   \Text(10,90)[c]{$\mu$}
   \LongArrow(40,68)(20,80)
   \Text(25,69)[c]{$k_1$}
   \Photon(80,90)(50,70){3}{3.5}
   \Text(90,90)[c]{$\nu$}
   \LongArrow(60,68)(80,80)
   \Text(75,69)[c]{$k_2$}
   \DashLine(20,45)(50,70){3}
   \DashLine(80,45)(50,70){3}
   \Text(50,28)[c]{$2iv_\chi^{-2}\bigl(2\mu_\chi^2\eta^{\mu\nu}+$}
   \Text(50,10)[c]{$k_1\cdot k_2\eta^{\mu\nu}
   -k_1^\nu k_2^\mu\bigr)$}
  \end{picture}
  \begin{picture}(100,100)(0,0)
   \Photon(30,90)(50,70){-3}{2.5}
   \Text(20,90)[c]{$\mu$}
   \Photon(70,90)(50,70){3}{2.5}
   \Text(80,90)[c]{$\nu$}
   \Photon(30,50)(50,70){3}{2.5}
   \Text(20,50)[c]{$\lambda$}
   \Photon(70,50)(50,70){-3}{2.5}
   \Text(80,50)[c]{$\sigma$}
   \Text(50,28)[c]{
   $-iv_\chi^{-2}\mu_\chi^2\bigl(\eta^{\mu\nu}\eta^{\lambda\sigma}+$}
   \Text(50,10)[c]{
   $\eta^{\mu\lambda}\eta^{\nu\sigma}+
   \eta^{\mu\sigma}\eta^{\nu\lambda}\bigr)$}
  \end{picture}
  \begin{picture}(100,100)(0,0)
   \Photon(30,90)(50,70){-3}{2.5}
   \Text(20,90)[c]{$\mu$}
   \Photon(70,90)(50,70){3}{2.5}
   \Text(80,90)[c]{$\nu$}
   \Gluon(30,50)(50,70){3}{2.5}
   \Text(20,50)[c]{$\lambda$}
   \Gluon(70,50)(50,70){-3}{2.5}
   \Text(80,50)[c]{$\sigma$}
   \Text(50,20)[c]{
   $-iv_\chi^{-2}\mu_\chi^2\eta^{\mu\nu}\eta^{\lambda\sigma}$}
  \end{picture}

  \end{center}

  \caption{\label{fig_feynmanrules}
  The Feynman rules. The dashed lines represent the dilaton $\chi$. For 
the sake of brevity, we omit the ordinary gauge-theoretical couplings 
and the non-Abelian self-interaction terms.
  The colour factors (i.e., $\delta^{\mathrm{AB}}$) are omitted.
  The last diagram applies to any combination of two non-identical gauge 
bosons, such as two sets of gluons of different colour.}
 \end{figure}
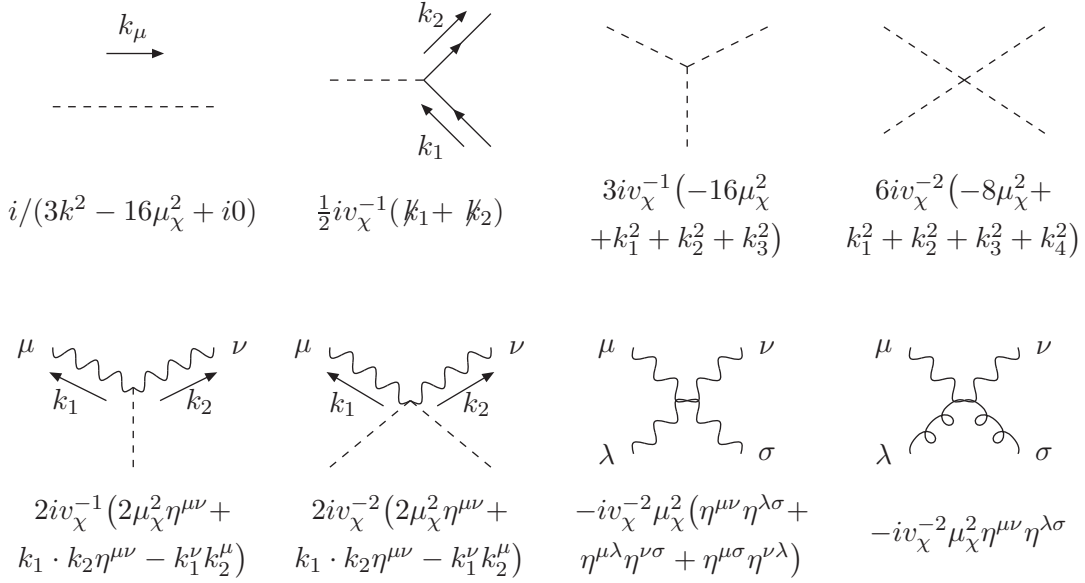

  It is a trivial matter to write down the Feynman rules corresponding 
to our Lagrangian 
$\mathcal{L}=\mathcal{L}_\mathrm{f}+\mathcal{L}_\mathrm{b}$.
  These are shown in figure~\ref{fig_feynmanrules}.

 \section{Tadpole cancellation in gauge-boson self-energy}
 \label{sec_gauge_tad}

  If gauge bosons are Goldstone bosons, their masses in the form of the 
anomalous tadpoles, must vanish. In our framework, the only 
contributions that can counteract the usual gauge theory contributions 
are due to the dilatonic contributions which are universal. It follows 
that the gauge theoretical tadpoles themselves must be universal at a 
scale $\Lambda_\mathrm{u}$ even though the gauge theoretical couplings 
themselves do not unify.

  Before proceeding to calculate the parameters of the theory, let us 
verify phenomenologically that this statement is indeed true.

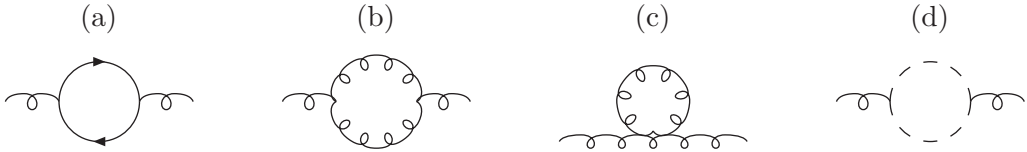
\begin{figure}[ht]
  \centerline{
  \begin{picture}(100,70)(0,0)
   \Text(50,62)[t]{(a)}
   \Gluon(15,25)(35,25){3}{1.5}
   \ArrowArcn(50,25)(15,0,180)
   \ArrowArcn(50,25)(15,180,360)
   \Gluon(65,25)(85,25){3}{1.5}
  \end{picture}
  \begin{picture}(100,70)(0,0)
   \Text(50,62)[t]{(b)}
   \Gluon(15,25)(35,25){3}{1.5}
   \GlueArc(50,25)(15,0,180){2.5}{4}
   \GlueArc(50,25)(15,180,360){2.5}{4}
   \Gluon(65,25)(85,25){3}{1.5}
  \end{picture}
  \begin{picture}(100,70)(0,0)
   \Text(50,62)[t]{(c)}
   \Gluon(15,10)(85,10){2.5}{5.5}
   \GlueArc(50,25)(11,-90,270){2.5}{6}
  \end{picture}
  \begin{picture}(100,70)(0,0)
   \Text(50,62)[t]{(d)}
   \Gluon(15,25)(35,25){3}{1.5}
   \DashCArc(50,25)(15,0,180){5}
   \DashCArc(50,25)(15,180,360){5}
   \Gluon(65,25)(85,25){3}{1.5}
  \end{picture}
  }
  \caption{\label{fig_gauge_boson_self_energy_ordinary}
  The usual gauge-theoretical contributions to the vacuum polarization 
operator.
  Diagram d represents the ghost contribution.
  There are, for the case of broken symmetry, also the Higgs and 
Goldstone-boson contributions which are not explicitly shown.
  }
 \end{figure}

  The tadpoles arise from diagrams which are shown in 
figure~\ref{fig_gauge_boson_self_energy_ordinary} for the case of QCD.

  This is a standard textbook calculation, but we have not found the 
results explicitly written out in the standard textbooks.
  We obtain
 \begin{equation}
  \Pi^\mathrm{tadpole}_{\mu\nu}=\frac{\alpha\Lambda^2}{8\pi}
  \eta_{\mu\nu}\left(8T_\mathrm{R}n_\mathrm{g}-2C_\mathrm{A}+
  n_\mathrm{Higgs}\right),
  \label{eqn_gaugeboson_tadpoles_ordinary}
 \end{equation}
  for SU(2)$_\mathrm{L}$ and QCD. $n_\mathrm{g}=3$ is the number of 
generations. $n_\mathrm{Higgs}$ is 1 for SU(2)$_\mathrm{L}$ (and 
U(1)$_\mathrm{Y}$), and 0 for QCD.
  $T_\mathrm{R}=1/2$, and $C_\mathrm{A}$ is 2 for SU(2)$_\mathrm{L}$
and 3 for QCD.
  $\Lambda^2$ is the UV cutoff of space-like $Q^2$ integration.
  A straightforward generalization of this formula makes it applicable 
also to the case of U(1)$_\mathrm{Y}$.

  The factor inside brackets is evaluated to be $6$ for QCD, $9$ for 
SU(2)$_\mathrm{L}$ and $21$ for U(1)$_\mathrm{Y}$ without the 
conventional factor $5/3$.
  Let us denote this as follows:
 \begin{equation}
  \Pi^\mathrm{tadpole}_{\mu\nu}=c_a\alpha_a\Lambda^2\eta_{\mu\nu},
  \qquad c_a=\frac{(21,9,6)}{8\pi}.
  \label{eqn_gaugeboson_tadpoles_ordinary_notation}
 \end{equation}

\begin{figure}[ht]
  \centerline{
   \includegraphics[width=12cm]{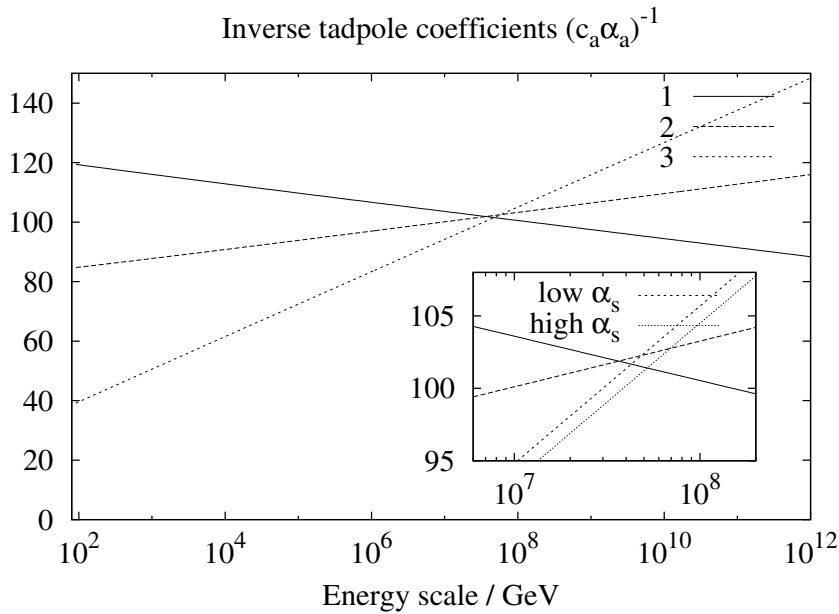}
  }
  \caption{\label{fig_gauge_tadpoles_plot}
  The inverse, $(c_a\alpha_a(\mu))^{-1}$ of the three gauge-theory 
tadpole coefficients as a function of the energy scale $\mu$.
  $c_a$ are dressed by one-loop anomalous dimensions as explained in the 
text.
  The main figure is for central $\alpha_\mathrm{s}(M_Z)=0.1176$. The 
inlay shows the unification region, for 
$\alpha_\mathrm{s}(M_Z)=0.1176\pm0.0020$ \cite{pdg}.
  The three-loop beta-function coefficients follow 
ref.~\cite{arasonetal}.}
 \end{figure}

  In figure~\ref{fig_gauge_tadpoles_plot}, we show the inverse tadpole 
coefficients $(c_a\alpha_a)^{-1}$ as a function of the energy scale.
  We see that the three tadpoles exhibit unification at a good level.
  $c_1\alpha_1$ and $c_2\alpha_2$ unify at
 \begin{equation}
  \Lambda_\mathrm{u}=3.6\times10^7~\mathrm{GeV},
  \label{eqn_unification_scale_phenomenological}
 \end{equation}
  when
 \begin{equation}
  (c_a\alpha_a)^{-1}=101.9.
  \label{eqn_tadpole_phenomenological}
 \end{equation}
  The error is small compared with the error in our prediction, with 
which we shall make comparison.

  We would like to make a number of remarks on the numbers shown in 
figure~\ref{fig_gauge_tadpoles_plot}.

  First, the couplings are calculated at the three-loop order, with the 
beta-function coefficients of ref.~\cite{arasonetal}.
  Second, the value $\alpha_\mathrm{s}(M_Z)=0.1176\pm0.0020$ quoted in 
figure~\ref{fig_gauge_tadpoles_plot} is relatively old \cite{pdg}.
  There is more statistical weight on the $e^+e^-$ data in these numbers 
as compared with the more modern numbers \cite{pdgnew} which place more 
weight on the $\tau$ decay data, and so we would like to think that the 
former is a more conservative estimate of $\alpha_\mathrm{s}(M_Z)$ as it 
applies to physics at $M_Z$.

  Finally, the values of $c_a$ used in 
figure~\ref{fig_gauge_tadpoles_plot} are not the leading order 
(one-loop) values which are discussed above.
  We have dressed $c_a$ by including, partially, the main 
next-to-leading order (two-loop) effect which is due to the anomalous 
dimensions.

  Fermionic loop contributions are modified by factor 
$(1-\gamma_\mathrm{f})$, where
 \begin{equation}
  \gamma_\mathrm{f}=\frac{\partial\ln Z(q^2)}{\partial\ln q^2}.
 \end{equation}
  $Z$ is the renormalization factor.
  The QCD contribution, which is appropriate for quarks, is 
$\gamma_\mathrm{f}=\alpha_\mathrm{s}/3\pi$ at the leading order.
  Gauge-boson loop contributions are modified by factor 
$(1-\gamma_\mathrm{v})$, where
 \begin{equation}
  \gamma_\mathrm{v}=\frac{\partial\ln\alpha}{\partial\ln q^2},
 \end{equation}
  and this is equal to $b_0\alpha$ at the leading order, where $b_0$ is 
the first beta-function coefficient.

  The values of $c_a$ as shown in figure~\ref{fig_gauge_tadpoles_plot} 
are dressed by including the $\gamma_\mathrm{v}$ factor and the QCD part 
of the $\gamma_\mathrm{f}$ factor.
  This is a small effect, but helps realize the unification of the 
tadpole coefficients.

 %

 \section{Derivation of the parameters}
 \label{sec_derivation}

 \subsection{General remarks}

  Let us now proceed to calculate the parameters of the theory.

  There are 5 unknown parameters, which are $v_\chi$, $\mu_\chi$ and the 
three gauge couplings $\alpha_a$.
  There is one additional parameter governing the UV running of 
$v_\chi$, but let us ignore this for now.

  In ordinary instances of dynamical symmetry breaking 
\cite{gribovewsb,odagirimagnetism,dasodagiri}, we would expect that 
$v_\chi$ is calculated from the self-energy integral of the Goldstone 
bosons, $\mu_\chi$ is determined by the self-energy of the Higgs bosons, 
and there is an additional constraint from the cancellation of the Higgs 
boson tadpole. Here the situation is similar, but certain modifications 
are necessary, namely
 \begin{enumerate}
  \item $v_\chi$ is determined by the dilaton self-energy rather than 
the gauge boson self-energy.
  \item Gauge boson self-energy yields constraints on $\alpha$.
  \item The dilatonic tadpole does not vanish by itself. We require an 
additional mass-generation mechanism.
  In the minimal framework, EWSB is necessary in order that the gauge 
bosons exist at all.
  \item $\mu_\chi$ does not arise directly from the dilatonic 
self-energy.
  It is calculated from the tadpole anomaly of the gauge bosons.
 \end{enumerate}

  Let us see how this works in practice.

 \subsection{Dilaton self-energy}

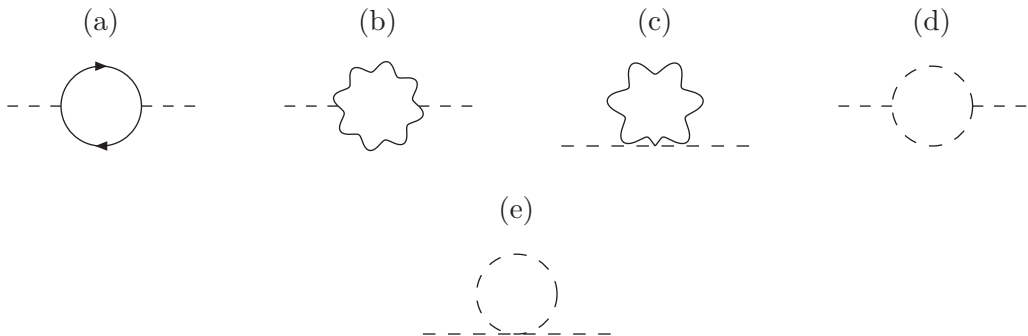
\begin{figure}[ht]

  \begin{center}

  \begin{picture}(100,70)(0,0)
   \Text(50,62)[t]{(a)}
   \DashLine(15,25)(35,25){5}
   \ArrowArcn(50,25)(15,0,180)
   \ArrowArcn(50,25)(15,180,360)
   \DashLine(65,25)(85,25){5}
  \end{picture}
  \begin{picture}(100,70)(0,0)
   \Text(50,62)[t]{(b)}
   \DashLine(15,25)(35,25){5}
   \PhotonArc(50,25)(15,0,180){-2}{4}
   \PhotonArc(50,25)(15,180,360){-2}{4}
   \DashLine(65,25)(85,25){5}
  \end{picture}
  \begin{picture}(100,70)(0,0)
   \Text(50,62)[t]{(c)}
   \DashLine(15,10)(85,10){5}
   \PhotonArc(50,25)(15,-90,270){-3}{6.5}
  \end{picture}
  \begin{picture}(100,70)(0,0)
   \Text(50,62)[t]{(d)}
   \DashLine(15,25)(35,25){5}
   \DashCArc(50,25)(15,0,360){5}
   \DashLine(65,25)(85,25){5}
  \end{picture}

  \begin{picture}(100,70)(0,0)
   \Text(50,62)[t]{(e)}
   \DashLine(15,10)(85,10){5}
   \DashCArc(50,25)(15,0,360){5}
  \end{picture}

  \end{center}

  \caption{\label{fig_dilaton_self_energy}
  Dilaton self-energy.}
 \end{figure}

  First, let us consider dilatonic self-energy.
  The contributions to $\Sigma$ are as shown in 
figure~\ref{fig_dilaton_self_energy}.

  The contributions corresponding to the five diagrams are divergent in 
general. The divergences are of three forms:
 \begin{eqnarray}
  \int\frac{d^4k}{(2\pi)^4}&=&0,\\
  \int\frac{d^4k}{(2\pi)^4}\frac1{k^2+i0}&=&
  -\frac{\Lambda^2}{16\pi^2},\\
  \int\frac{d^4k}{(2\pi)^4}\frac1{k^2+i0}\frac1{(k-q)^2+i0}&=&
  \frac1{16\pi^2}\ln\frac{\Lambda^2}{-q^2}.
 \end{eqnarray}
  The quartic divergence contains no poles, and so we set it to zero.

  The fermionic contribution of figure~\ref{fig_dilaton_self_energy}a is 
evaluated to be
 \begin{equation}
  \Sigma_\mathrm{(a)}=\frac{\#_\mathrm{f}\Lambda^2q^2}{32\pi^2v_\chi^2},
 \end{equation}
  where $\#_\mathrm{f}=24$ is the number of fermionic degrees of 
freedom. We have omitted a $q^4$ term with a finite coefficient, as we 
shall be introducing a counterterm later on.

  The contribution of figure~\ref{fig_dilaton_self_energy}b is evaluated 
to be
 \begin{eqnarray}
  \Sigma_\mathrm{(b)}&=&
  -\frac{\#_\mathrm{v}q^2}{16\pi^2v_\chi^2}
  \left(\Lambda^2+q^2\ln\frac{\Lambda^2}{-q^2}\right)
  -\frac{3\#_\mathrm{v}\mu_\chi^2}{4\pi^2v_\chi^2}
  \left(2\Lambda^2+q^2\ln\frac{\Lambda^2}{-q^2}\right)
  \nonumber\\
  &&-\frac{2\#_\mathrm{v}\mu_\chi^4}{\pi^2v_\chi^2}
  \ln\frac{\Lambda^2}{-q^2}.
 \end{eqnarray}
  where $\#_\mathrm{v}=12$ is the number of vector bosons.
  The last term is gauge dependent, and we have used the Feynman gauge. 
  However, we shall not be discussing the $\mu_\chi^4$ term for the rest 
of this study.
  Again we have omitted a $q^4$ term.

  The contribution of figure~\ref{fig_dilaton_self_energy}c is evaluated 
to be
 \begin{equation}
  \Sigma_\mathrm{(c)}=
  \frac{\#_\mathrm{v}\mu_\chi^2\Lambda^2}{2\pi^2v_\chi^2}.
 \end{equation}
  This is in the Feynman gauge.

  In the sum of contributions $\Sigma_\mathrm{(a-c)}$, we notice that 
the anomalous $\Lambda^2q^2$ terms cancel, provided 
$2\#_\mathrm{v}=\#_\mathrm{f}$.
  In our world, assuming that the neutrinos are Dirac particles, this 
requires that the number of generations is equal to three.
  It is easy to verify that $\Lambda^2q^2$ terms which are present in 
$\Sigma_\mathrm{(d)}$ and $\Sigma_\mathrm{(e)}$ mutually cancel.

  The $\Lambda^2\mu_\chi^2$ terms are also anomalous, but we cannot see 
how they will cancel.
  On the other hand, $\Sigma$ by itself does not contain a 
mass-generation mechanism. That is, $\mu_\chi^2$ remains zero if it is 
zero to start off with.
  Thus the mass term needs to be inserted by other means, and when this 
is done, the resultant term must be equal to $16\mu_\chi^2$. Thus we 
think it reasonable to drop $\Lambda^2\mu_\chi^2$ term as being 
unphysical.
  The terms that are physically significant are therefore
 \begin{equation}
  \Sigma_\mathrm{(a-c)}=-\frac{\#_\mathrm{v}q^4}{16\pi^2v_\chi^2}
  \ln\frac{\Lambda^2}{-q^2}-
  \frac{3\#_\mathrm{v}\mu_\chi^2q^2}{4\pi^2v_\chi^2}
  \ln\frac{\Lambda^2}{-q^2}.
  \label{eqn_dilaton_self_energy_abc}
 \end{equation}

  The analogous contributions from 
figures~\ref{fig_dilaton_self_energy}d, \ref{fig_dilaton_self_energy}e 
are evaluated to be
 \begin{equation}
  \Sigma_\mathrm{(d,e)}=-\frac{q^4}{32\pi^2v_\chi^2}
  \ln\frac{\Lambda^2}{-q^2}+
  \frac{\mu_\chi^2q^2}{3\pi^2v_\chi^2}
  \ln\frac{\Lambda^2}{-q^2},
  \label{eqn_dilaton_self_energy_de}
 \end{equation}
  when $q^2\gg M_\chi^2$.
  For $\#_\mathrm{v}=12$, the latter contributions are one order of 
magnitude smaller.
  Furthermore, they will be much more suppressed if the calculation is 
done self-consistently, because both $\mu_\chi^2$ and $v_\chi^{-2}$, 
which become functions of the internal momenta, decay rapidly at high 
energies.
  Let us therefore neglect eqn.~(\ref{eqn_dilaton_self_energy_de}).

  Self-consistency requires that $\Sigma=-3q^2$.
  Note that $\mu_\chi$ will be fixed by gauge-boson tadpole-cancellation 
conditions.
  This implies
 \begin{equation}
  v^2_\chi(q^2)=\frac{q^2}{4\pi^2}\ln\frac{\Lambda^2}{-q^2}+
  \frac{3\mu_\chi^2}{\pi^2}\ln\frac{\Lambda^2}{-q^2}.
 \end{equation}
  The first term is problematic.
  If $\Lambda$ in both terms are the same, this will induce a tachyonic 
pole at $q^2=-12\mu_\chi^2$, and $v^2_\chi$ will be negative for large 
and space-like $q^2<-12\mu_\chi^2$.
  Since it runs, we cannot subtract it away at all energy scales.
  The best that can be done is presumably to subtract it away at the 
symmetry-breaking scale.
  This implies that $\Lambda$ will be replaced by $\Lambda_\mathrm{u}$ 
in the first term, whereas $\Lambda$ in the second term remains the UV 
cutoff. Our final expression for $v_\chi$ reads
 \begin{equation}
  v^2_\chi(q^2)=\frac{-q^2}{4\pi^2}\ln\frac{-q^2}{\Lambda_\mathrm{u}^2}+
  \frac{3\mu_\chi^2(q^2)}{\pi^2}\ln\frac{\Lambda_\mathrm{cut}^2}{-q^2}.
  \label{eqn_running_v}
 \end{equation}
  The second term is the dominant contribution when discussing symmetry 
breaking, and our results will be derived solely from it.
  The first term is necessary, nevertheless, to protect both $v_\chi^2$ 
and $\mu_\chi^2$ from growing negative at high energies.

 \subsection{Gauge-boson self-energy}

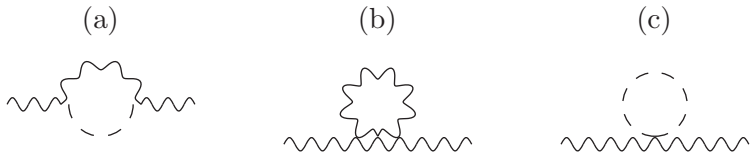
\begin{figure}

  \begin{center}

  \begin{picture}(100,70)(0,0)
   \Text(50,62)[t]{(a)}
   \Photon(15,25)(35,25){2.5}{2.5}
   \Photon(65,25)(85,25){2.5}{2.5}
   \PhotonArc(50,25)(15,0,180){-2}{4.5}
   \DashCArc(50,25)(12,180,360){5}
  \end{picture}
  \begin{picture}(100,70)(0,0)
   \Text(50,62)[t]{(b)}
   \Photon(15,10)(85,10){2.5}{8.5}
   \PhotonArc(50,25)(12,-90,270){-2.5}{8.5}
  \end{picture}
  \begin{picture}(100,70)(0,0)
   \Text(50,62)[t]{(c)}
   \Photon(15,10)(85,10){2.5}{8.5}
   \DashCArc(50,25)(12,-90,270){5}
  \end{picture}

  \end{center}

  \caption{\label{fig_gauge_boson_self_energy_new}
  New contributions to the vacuum polarization operator.
  }
 \end{figure}

  The gauge boson tadpoles, which were calculated in 
section~\ref{sec_gauge_tad}, need to be cancelled by the new 
contributions of the form shown in 
figure~\ref{fig_gauge_boson_self_energy_new}.

  These contributions are easy to calculate, and we obtain
 \begin{equation}
  \Pi_{\mu\nu}^\mathrm{tadpole}=-\frac{1}{8\pi^2}\eta_{\mu\nu}
  \int dQ^2\frac{\mu_\chi^2(Q^2)}{v_\chi^2(Q^2)}
  \left(\#_\mathrm{v}+\frac1{16\mu_\chi^2/Q^2+3}\right).
 \end{equation}
  It is essential that we use the running $\mu_\chi$ and $v_\chi$.

  We now compare this with the result of section~\ref{sec_gauge_tad}.
  There should be cancellation at all scales, and so we obtain
 \begin{equation}
  c_i\alpha_i(Q^2)=\frac{\mu^2_\chi(Q^2)}{8\pi^2v_\chi^2(Q^2)}
  \left(\#_\mathrm{v}+\frac1{16\mu_\chi^2/Q^2+3}\right).
  \label{eqn_UV_running_coupling}
 \end{equation}
  Obviously this will only hold when $c_i\alpha_i$ are approximately 
universal.
  We then substitute the second term of eqn.~(\ref{eqn_running_v}) to 
obtain the following prediction:
 \begin{equation}
  c_i\alpha_i(\Lambda_\mathrm{u}^2)=
  \frac1{24\ln(\Lambda_\mathrm{cut}^2/\Lambda_\mathrm{u}^2)}
  \left(\#_\mathrm{v}+
  \frac1{16\mu_\chi^2/\Lambda_\mathrm{u}^2+3}\right).
  \label{eqn_coupling_prediction}
 \end{equation}
  By omitting the second term, we obtain the predictions that are quoted 
in the introduction.
  By adopting $\Lambda_\mathrm{cut}=\widetilde{M}_\mathrm{Pl}$ and 
eqn.~(\ref{eqn_unification_scale_phenomenological}) for 
$\Lambda_\mathrm{u}$, we obtain the remarkably accurate result
$(c_i\alpha_i)^{-1}=100$.

 \subsection{The behaviour of the dilatonic mass}
 \label{sec_dilatonic_mass_behaviour}

  We would now like to start the discussion of what causes the 
spontaneous breaking of gauge--dilatation symmetry.
  First of all, we need to know what physics generates $\mu_\chi^2$.

\begin{figure}[ht]
  \centerline{
  \begin{picture}(100,70)(0,0)
   \ArrowArcn(50,25)(15,0,180)
   \ArrowArcn(50,25)(15,180,270)
   \ArrowArcn(50,25)(15,270,360)
   \Photon(15,25)(35,25){3}{1.5}
   \Photon(65,25)(85,25){3}{1.5}
   \DashLine(50,40)(50,60){5}
  \end{picture}
  }
  \caption{\label{fig_three_point_anomaly}
  The one-loop anomalous amplitude which generates $\mu_\chi^2$.}
 \end{figure}
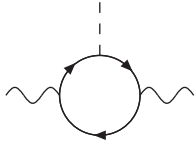

  Consider the $\mu_\chi^2/v_\chi$ interaction term for the 
dilaton--gauge-boson--gauge-boson vertex that is shown at the bottom 
left of figure~\ref{fig_feynmanrules}.
  The presence of a tadpole anomaly in the gauge-boson propagator 
implies that this interaction term arises automatically in amplitudes 
such as that shown in figure~\ref{fig_three_point_anomaly}.

  The interaction of the dilaton, at least when it is soft, has the form 
of the inverse propagator.
  The insertion of a soft dilaton line in 
figure~\ref{fig_three_point_anomaly} therefore removes a fermion line.
  The result of this procedure is that the induced three-point function 
will be proportional to the anomalous tadpole.
  We obtain
 \begin{equation}
  \frac{4\mu_\chi^2(Q^2)}{v_\chi(Q^2)}=
  \int_{Q^2} dQ^{\prime2}\frac{2c_i\alpha_i(Q^{\prime2})}{v_\chi(Q^2)},
 \end{equation}
  so that 
 \begin{equation}
  \frac{d\mu_\chi^2(Q^2)}{dQ^2}=-
  \frac12c_i\alpha_i(Q^2).
  \label{eqn_mu_ode}
 \end{equation}
  Thus $\mu_\chi^2$ decays at high energy.

  This procedure gives us an answer to the problem of gauge invariance.
  The anomalous tadpole arises from amplitudes such as
 \begin{equation}
  \int d^4k \frac{\alpha}{4\pi^3i}
  \frac1{k^2+i0}\frac1{(k-q)^2+i0}
  \mathrm{Tr}
  \left[\not\! k\gamma_\mu(\not\! k-\not\! q)\gamma_\nu\right].
 \end{equation}
  The contraction with $q^\nu$ yields
 \begin{equation}
  \int d^4k \frac{\alpha}{i\pi^3}\left[
  \frac{(k-q)_\mu}{(k-q)^2+i0}-\frac{k_\mu}{k^2+i0}
  \right].
  \label{eqn_ward_tadpole}
 \end{equation}
  Provided that $\alpha$ decays sufficiently fast at large $k^2$, we can 
replace $k-q$ in the first term with $k$, and the resultant integral 
will be zero.
  That is, the mass term will be of the form that kills longitudinal 
contributions, and therefore the practical recipe will be that we can 
use the Feynman gauge in our calculations.
  Note that if $\alpha$ does not decay, eqn.~(\ref{eqn_ward_tadpole}) 
will yield $\Lambda^2q_\mu$. We shall now show that $\alpha(Q^2)$ decays 
faster than $1/Q^2$.

  We substitute the first term of eqn.~(\ref{eqn_UV_running_coupling}) 
in eqn.~(\ref{eqn_mu_ode}):
 \begin{equation}
  \frac{d\mu_\chi^2(Q^2)}{dQ^2}=
  -\frac{3\mu_\chi^2(Q^2)}{4\pi^2v_\chi^2(Q^2)}.
 \end{equation}
  This, together with eqn.~(\ref{eqn_running_v}) allows us to obtain the 
running $\mu_\chi^2$, $v_\chi^2$ and $\alpha$ at above 
$\Lambda_\mathrm{u}$ scale.

  The running is at first dominated by the second term of 
eqn.~(\ref{eqn_running_v}).
  In this region, we have
 \begin{equation}
  \mu_\chi^2(Q^2)=
  -c_i\alpha_i(\Lambda_\mathrm{u}^2)\frac{Q^2}2+\mbox{const}.
 \end{equation}
  However, as $\mu_\chi^2$ reaches zero, the first term of 
eqn.~(\ref{eqn_running_v}) starts to dominate, causing $\mu_\chi^2$ to 
fall as a power.
  It should be noted that $v_\chi^2(Q^2)$ decreases with scale at first, 
after which it increases proportionally to $Q^2$.
  The phase transition is not quite second order in the sense that 
$v_\chi$ does not vanish at $\Lambda_\mathrm{cut}$.
  The ratio of $\mu_\chi^2$ and $v_\chi^2$, on the other hand, decreases 
monotonically, and hence so does $\alpha$.

  Since $\mu_\chi^2$ is already small when $v_\chi^2$ starts to rise 
quadratically, a useful approximation would be
 \begin{equation}
  \mu_\chi^2(Q^2)=\mbox{max}\left(\mu_\chi^2(\Lambda_\mathrm{u}^2)-
  c_i\alpha_i(\Lambda_\mathrm{u}^2)
  \frac{Q^2-\Lambda_\mathrm{u}^2}2,0\right).
  \label{eqn_running_mu_approximate}
 \end{equation}

 \subsection{EWSB as the origin of scales}

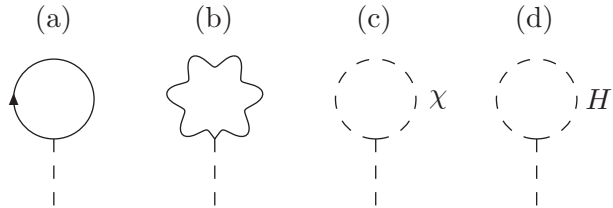
\begin{figure}[ht]
  \centerline{
  \begin{picture}(230,110)(0,0)
   \Text(20,95)[c]{(a)}
   \DashLine(20,25)(20,50){5}
   \ArrowArcn(20,65)(15,360,0)
   \Text(80,95)[c]{(b)}
   \DashLine(80,25)(80,50){5}
   \PhotonArc(80,65)(15,-90,270){-3}{6.5}
   \Text(140,95)[c]{(c)}
   \Text(164,65)[c]{$\chi$}
   \DashLine(140,25)(140,50){5}
   \DashCArc(140,65)(15,-90,270){5}
   \Text(200,95)[c]{(d)}
   \Text(224,65)[c]{$H$}
   \DashLine(200,25)(200,50){5}
   \DashCArc(200,65)(15,-90,270){5}
  \end{picture}
  }
  \caption{\label{fig_tadpole_diagrams}
  The three contributions to the dilatonic tadpole (a--c) and the Higgs 
boson loop (d) which would have cancelled the massive contributions
of diagrams a, b.
  }
 \end{figure}

  We now consider the tadpole cancellation condition.
  The relevant diagrams are shown in 
figure~\ref{fig_tadpole_diagrams}a--c.
  The Higgs boson should not couple to the dilaton directly as it breaks 
both gauge and scaling symmetries, and so the contribution of 
figure~\ref{fig_tadpole_diagrams}d is zero.

  One would naively expect that, provided that we can neglect the masses 
of the fermions, the contribution of diagram a is zero, and diagrams b 
and c must mutually cancel.
  However, this cannot work since diagram b will at best give a 
contribution that has the same sign as that of diagram c:
 \begin{equation}
  \mathcal{A}_\mathrm{(b)}=
  \frac{\#_{\mathrm{v}}\mu_\chi^2}{2\pi^2v_\chi}
  \int dQ^2,
  \label{eqn_tadpole_b}
 \end{equation}
 \begin{equation}
  \mathcal{A}_\mathrm{(c)}=\int
  \frac{\mu_\chi^2(Q^2)}{2\pi^2v_\chi(Q^2)}
  \frac{dQ^2}{16\mu_\chi(Q^2)^2/Q^2+3}.
  \label{eqn_tadpole_c}
 \end{equation}

  We conclude that somehow the masses of fermions, gauge bosons and the 
Higgs boson has a part to play.
  But this seems unintuitive, since eqn.~(\ref{eqn_tadpole_b}) yields a 
contribution on the order of $\mu_\chi^2\Lambda^2/v_\chi$, whereas the 
contribution of figure~\ref{fig_tadpole_diagrams}a is of the order of 
$m_t^2\Lambda^2/v_\chi$, and is five orders of magnitude smaller.

  The answer is that the $\mu_\chi^2/v_\chi$ coupling of 
figure~\ref{fig_tadpole_diagrams}b is induced through an anomaly as 
discussed in the previous section.
  If we evaluate $\mathcal{A}_\mathrm{(b)}$ instead as an all-order 
quantity using a Dyson--Schwinger formalism (bare vertex, dressed 
propagator), we will obtain zero so long as the gauge bosons are 
massless.
  On the other hand, the dilaton is massive, so the Dyson--Schwinger 
approach (massless vertex, massive propagator) to 
$\mathcal{A}_\mathrm{(c)}$ yields
 \begin{equation}
  \mathcal{A}_\mathrm{(c)}^\mathrm{DS}=
  -\int\frac{\mu_\chi^2(Q^2)}{\pi^2v_\chi(Q^2)}
  \frac{dQ^2}{16\mu_\chi^2(Q^2)/Q^2+3}.
  \label{eqn_tadpole_c_ds}
 \end{equation}
  Note that because $\mu_\chi^2(Q^2)$ is significant only near the 
$\Lambda_\mathrm{u}$ scale, this integral will be on the order of 
$\Lambda_\mathrm{u}^4/v_\chi$.

  Now let us consider the massive contribution to fermion and boson 
loops.
  Provided that all of these masses are generated by the Higgs 
mechanism, the contributions will necessarily have the form of 
Higgs--dilaton mixing.
  If the Higgs-boson itself has a vanishing tadpole, these will all 
vanish, except for the self-coupling contribution which, had it existed, 
will have the form of figure~\ref{fig_tadpole_diagrams}d.
  This last contribution will be given by
 \begin{equation}
  -\mathcal{A}_\mathrm{(d)}=-
  \frac3{32\pi^2v_\chi}\int M_H^2(Q^2)dQ^2.
  \label{eqn_tadpole_d_higgs}
 \end{equation}
  $M_H$ is the mass of the Higgs boson.

  Let us assume that EWSB is caused by dynamical symmetry breaking 
originating at the same cutoff scale $\Lambda_\mathrm{cut}$ as 
gauge--dilatation symmetry breaking.
  This is natural, because if the latter symmetry breaking requires the 
former symmetry breaking, the former symmetry breaking will be forced to 
occur even in the absence of an interaction which grows strong at some 
large energy scale. This is when the symmetry-broken vacuum is more 
energetically favourable.

  Equation~(\ref{eqn_tadpole_d_higgs}) is dominated by the region near 
$\Lambda_\mathrm{cut}$.
  Using the dynamical symmetry breaking hypothesis, it is easy to 
estimate $M_H^2(Q^2)$ near $\Lambda_\mathrm{cut}$.
  We use the formalism of ref.~\cite{gribovewsb,dasodagiri}.
  In the high-energy limit, eqn.~(26) of ref.~\cite{dasodagiri} is 
approximated by eqn.~(44) of ref.~\cite{gribovewsb} and we obtain
 \begin{equation}
  M_H^2(Q^2)\approx\frac{3}{2v_H^2\pi^2}m_t^4(\Lambda_\mathrm{cut}^2)
  \ln(\Lambda_\mathrm{cut}^2/Q^2).
 \end{equation}
  $v_H=246.22$~GeV is the Higgs condensate.
  By eqn.~(15) of ref.~\cite{dasodagiri} we then obtain
 \begin{equation}
  M_H^2(Q^2)\approx\frac{3}{128v_H^2\pi^2}M_H^4
  \ln(\Lambda_\mathrm{cut}^2/Q^2).
 \end{equation}
  $M_H^4$ on the right-hand side refers to the low-energy value 
$M_H\approx120$~GeV.
  Substituting this in eqn.~(\ref{eqn_tadpole_d_higgs}) yields
 \begin{equation}
  \mathcal{A}_\mathrm{(d)}\approx
  \frac{9M_H^4\Lambda_\mathrm{cut}^2}{4096v_H^2v_\chi\pi^4}.
 \end{equation}

  The contribution of eqn.~(\ref{eqn_tadpole_c_ds}), on the other hand, 
is evaluated easily using the approximation of 
eqn.~(\ref{eqn_running_mu_approximate}). We obtain
 \begin{equation}
  \mathcal{A}_\mathrm{(c)}^\mathrm{DS}\approx-
  \frac{\mu_\chi^2(\Lambda_\mathrm{u}^2)}{3\pi^2v_\chi}\int
  \sqrt{1-Q^2c_i\alpha_i/2\mu_\chi^2(\Lambda_\mathrm{u}^2)}\ dQ^2=
  -\frac{4\mu_\chi^4(\Lambda_\mathrm{u}^2)}{9\pi^2c_i\alpha_iv_\chi}.
 \end{equation}
  Let us denote $M_\chi^2=16\mu_\chi^2(\Lambda_\mathrm{u})/3$.
  The tadpole cancellation condition then yields
 \begin{equation}
  M_\chi^4=\frac{9c_i\alpha_iM_H^4\Lambda_\mathrm{cut}^2}{64v_H^2\pi^2}.
 \end{equation}
  Using $\Lambda_\mathrm{cut}=\widetilde{M}_\mathrm{Pl}$ yields 
$M_\chi\approx1.3\times10^9$~GeV.

  Since we expect $M_\chi\approx\Lambda_\mathrm{u}$, this prediction is 
higher than what we had hoped for.
  We think it significant nevertheless, that a sensible value of 
$M_\chi$ does emerge from simple considerations of dynamical symmetry 
breaking.

 \section{Summary and outlook}
 \label{sec_conclusion}

 \subsection{Summary}

  The presence of a tadpole anomaly suggests that gauge bosons are not 
fundamental, and that they are Goldstone bosons of some symmetry.
  We have argued that if this is the case, the symmetry is that of 
the rotation between the gauge-theoretical phases and space-time 
coordinates. A dilatonic scalar particle behaves as the Higgs mode.

  A necessary condition for this programme is that the three 
gauge-theory tadpoles unify.
  We have verified this statement. The unification occurs at a rather 
low scale of $10^7$ to $10^8$~GeV.

  We calculated the parameters of this theory within the dynamical 
symmetry breaking picture, by imposing self-consistency conditions.
  The value of the unified coupling turns out to be given by the inverse 
of the logarithm of the two cutoff scales. Setting the upper cutoff 
equal to the reduced Planck mass yields a number which agrees with the 
phenomenological value to within a few \%.

  The origin of the large difference in the two scales can be explained 
as being due to EWSB. Without EWSB, there cannot be gauge--dilatation 
symmetry breaking.
  By considering the tadpole cancellation condition of the dilaton 
$\chi$, we predicted $M_\chi\approx10^9$~GeV. This is somewhat higher 
than the unification scale, and necessitates further investigation.

 \subsection{Theoretical outlook}

  Given that gauge--dilatation symmetry breaking cannot occur without 
EWSB, dynamical EWSB can proceed without the presence of a strong UV 
interaction, because one gains fermionic Casimir energy through 
gauge--dilatation symmetry breaking.
  As a topic for future study, it will be interesting to analyze how the 
scale hierarchy between EWSB scale and the Planck scale may arise, 
similarly to ref.~\cite{dasodagiri} but now including the $10^8$~GeV 
scale physics.

  Another possible topic for future study will be the application of 
similar ideas to gravitation.
  If there is a condition on tadpole cancellation that is similar to the 
present case, we would expect that there arises a light gravitational 
dilaton with a mass on the order of the electroweak scale. This will be 
a candidate for dark matter.

  In our work, the $\Lambda^2q^2$ anomaly in the dilatonic self-energy 
cancels provided that the number of fermionic generations equals three.
  In the case of gravitation, if similar ideas are applicable, we would 
like such anomalous terms to survive, so that the form factor will be 
large and produce the correct value of Newton's constant.

 \subsection{Phenomenology}

  The direct experimental confirmation of our proposal will be 
difficult.
  For example, the production of a $\chi$ resonance will require 
collisions at a centre-of-mass energy of about $10^7$ or $10^8$~GeV.

  There are a number of other possibilities:
 \begin{itemize}
  \item The dilatonic contribution to $\gamma\gamma\to\gamma\gamma$ 
elastic scattering is suppressed at the amplitude level by 
$Q^2/v_\chi^2$.
  The Standard Model contributions being suppressed by 
$\alpha_\mathrm{EM}^2$, the two contributions will become comparable at 
about $10^6$ or $10^7$~GeV.
  \item The dilaton is associated with conserved scaling symmetry, which 
is broken by the Higgs mechanism. In other words, the non-conserved 
trace part of the symmetry current $\sim T_{\mu\nu}$ is due to the Higgs 
boson, and this condition yields dilaton--Higgs mixing.
  The mixing angle is of order $v_HM_H^2/(v_\chi M_\chi^2)$, and is 
therefore tiny when considering low-energy phenomenology.
  \item As mentioned above, the EWSB-scale gravitational dilaton will be 
an indirect prediction of our theory but, in reflection, this prediction 
will hold true even without the presence of a gauge--dilatation symmetry 
breaking.
  The lack of such a particle will not rule out gauge--dilatation 
symmetry breaking either in any way.
  \item Cosmology will be affected since there will be no propagating 
fields above the $\Lambda_\mathrm{u}$ scale. Gravity will be a possible 
exception.
 \end{itemize}

 \acknowledgments

  The author thanks B.~R.~Webber, whose insightful remarks dispelled the 
confusion regarding cancellations that was present in an earlier version 
of this work.

  The author also thanks the members of his research group for their 
encouragement. 
  It would not have been possible to work on this topic without their 
generosity and support.

 \end{document}